\documentclass{elsart}

\usepackage{epsfig}
\usepackage{epsf}

\begin{document}

\begin{frontmatter}
\title{Anomalous Diffusion in Quasi One Dimensional Systems}
\author{F. M. Cucchietti},
\author{H. M. Pastawski}
\address{LANAIS de RMS, Facultad de Matem\'{a}tica, Astronom\'{i}a y F\'{i}sica, Universidad
Nacional de C\'{o}rdoba, C\'{o}rdoba, Argentina} 
\begin{abstract}
In order to perform quantum Hamiltonian dynamics minimizing localization effects, 
we introduce a quasi-one dimensional tight-binding model whose mean free path is 
smaller than the size of the sample. This size, in turn, is smaller than the localization 
length. We study the return probability to the starting layer using direct diagonalization of 
the Hamiltonian. We create a one dimensional excitation 
and observe sub-diffusive behavior for times larger than the Debye time but shorter 
than the Heisenberg time. The exponent corresponds to the fractal dimension $d^{*} \sim 0.72$ 
which is compared to that calculated from the eigenstates by means of the 
inverse participation number. 
\end{abstract}
\end{frontmatter}

The classical kinetic theory predicts that, in a disordered system, the
return probability of an excitation decays with a diffusive law $P(t)\sim
(4\pi Dt)^{-d/2}$, where $D$ is the diffusion coefficient and $d$ the
dimension of the system. This fact has been extensively used in many areas
of physics, in particular in electronic transport. However, the theory of
quantum localization\cite{AALR}, on the basis of steady state transport
properties, made clear that different regimes arise as a function of the
disorder $W$. For $W\rightarrow \infty $, the eigenstates are completely
localized and wave packets do not move through the system (insulating
phase). If $W\rightarrow 0$, the motion is completely ballistic with a
velocity $Ja/\hbar $. Between these two limits the diffusive behavior
(metallic phase) is observed. The particles move around freely between
collisions for a certain average length $\ell $, called the mean free path.

More recently, theoretical studies have considered the regime where the
critical amount of disorder $W_{C}$ is such that the system is at the
metal-insulator transition ($W_{C}$ $\sim zJ$ is the typical exchange
energies with $z$ neighbors at distance $a$). In that case, the dynamics of
the system is diffusive, but with a smaller exponent implying $d^{*}<d$ \cite
{Huckestein}. This reduction in the effective dimension is attributed to the
(multi)fractality of the eigenstates at the transition. On the other hand,
the diffusion of a spin excitation was directly observed \cite{McDowell} in
NMR\ experiments. When the spin network has a cubic structure filling the space, 
the intensity of the excitation decays as $(t-t_{0})^{-1.5}$,
which is the expected value for diffusion in a three dimensional system.
However, the same experiment performed in a chain-like structure (powdered
polyethylene) shows anomalous exponents in the diffusion of the excitation;
namely $0.9$ and $0.7$ for the crystalline and amorphous parts of the sample
respectively. In these systems, the role of disorder is played by typical
energy differences $W$ between states, which are smaller than exchange
energies $zJ.$ This assures\cite{Ander_Nobel} a diffusive dynamics with\cite
{QZeno} $D\sim Ja^{2}/\hbar $. Therefore, the effective dimensions $d^{*}=1.8
$ and $1.4$ corresponding to these values reflect the spatial structure of
the spin network.

In this work we study the conditions to reach the diffusive regime from
actual quantum dynamics, namely the numerical study of model Hamiltonians.
In particular, we developed a quasi one dimensional tight binding model\textit{\ }system,
which we called \textit{the Stars necklace model, }whose basic unit (layer)
is a highly connected cluster (see inset in figure 1) with $N$ sites and intralayer 
hopping $J=V/N^{0.5}$.
Disorder is introduced through on site energies characterized by a random
distribution of width $W$ . The initial wave function is a packet defined in
one layer of the system. We are interested in the probability of return to
the layer, $P$, which is the sum of probabilities of finding the particle in
every site of the initial layer. To calculate the dynamics of the system we
perform an exact diagonalization of the Hamiltonian. One key aspect of the
numerical calculation is that the mean free path $\ell $, the size of the
system $L$ and the localization length $\xi $ must obey some restrictive
relationships. The condition for a diffusive regime is $\ell $ much smaller
than $L$, hence assuring that the particle will collide many times before
reaching the boundaries of the system. In turn, to stay away from the
localized regime, $L$ must be smaller than $\xi $. For a strictly one
dimensional wire $\xi =2\ell $, while for strips and bars with a given
number of transverse modes (channels) $M$ the localization length is
expected to go\cite{MacKinnon} as $\xi =2M\ell $. In our model the $M=N-1$
channels available for transport have the \textit{same} group velocity $v=2Va/\hbar$.
This striking feature of the model allows to reach a one dimensional
diffusive dynamics when $N\rightarrow \infty .$ For finite $N$ it provides
an optimal representation for one-dimensional excitations. We studied many
system sizes and amounts of disorder, a typical evolution is shown in
figure 1. The particular system for this figure has $N=12$ and a perimeter of $%
100a$, $W=3V$, $\ell \sim 6a$ and $\xi \sim 140a$ . We see that after a
ballistic time $\ell /v$, the evolution follows a power law which indicates
a diffusive behavior. Nevertheless, the exponent of this power law is somewhat different
 from the expected one dimensional value $0.5$. As in the examples
mentioned above, this anomalous exponent could be due to a fractal effective
dimension of the system. The fitting of the evolution to a power law with
a free exponent resulted in an effective dimension of the system $%
d^{*}\sim 0.7$. In our model, a possible cause for a fractality in
the eigenfunctions of the system is disorder. Strongly localized stated in
the band tails are confined around some random points. This means that they
represent ``holes'' in the real space allowed to extended wave functions,
thus making the effective dimension of the system smaller than the real one.
For times longer than the ones showed in the figure, the autocorrelation
function saturates, this is a finite size effect (the saturation value
depends linearly on the system size). We also observed that a magnetic field
does not change the exponent in the power law noticeable, but reduces the
value of the saturation, meaning that there are fewer localized states.
Another way to study how the eigenstates of energy $\varepsilon $ are
occupying some fraction of the volume of space is by means of the inverse
participation number $p^{-1}=\sum_{i}\left| \varphi _{i}(\varepsilon
)\right| ^{4}$ \cite{MacKinnon}. For plane waves one obtains that $%
p=L^{d}$, i.e. equals the volume of the system. For a localized state $%
p$ is proportional to the volume in which the state has a non-vanishing
amplitude. However if the states are extended but fractal, in the thermodynamic
limit it diverges as $p=L^{d^{*}}$, with $d^{*}$ an effective dimension that may be
different from the dimension $d$ of the ordered system. We calculated the
inverse participation number for each of the eigenstates of the system and
through it the effective dimension ratio $d^{*}/d$ of everyone of them. The
results (shown in figure 2) are in very good agreement with the effective
dimension calculated through the fitting of the autocorrelation function
depicted in Fig. 1. Summarizing, we have introduced a numerical Hamiltonian
model whose exact solution shows a regime with sub-diffusive behavior.
Moreover, we presented hints of a fractal dimension of the extended
eigenstates induced by the presence of disorder. By hindering particles from
a fraction of the available real space, disorder induces a weak breaking of
the ergodicity that anticipates the non-ergodicity associated with full
localization.

\begin{figure}[tb]
\centering \leavevmode
\center{\epsfig{file=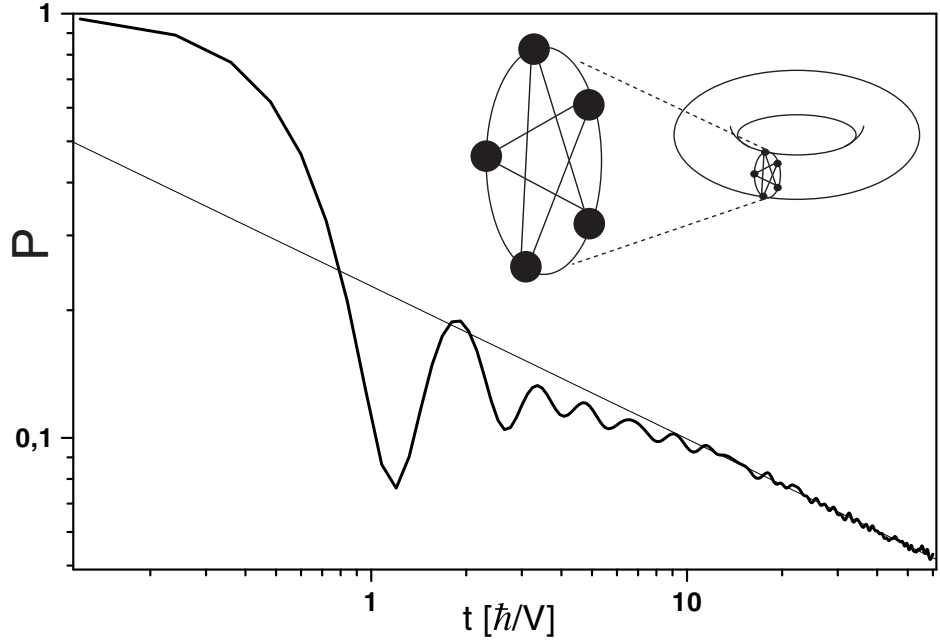, ,width=16cm,angle=270}}
\caption{Evolution of the layer autocorrelation function (in thick solid
line). The thin line is a best fit to $(4\pi Dt)^{d^{*}/2}$, with the
results $d^{*}=0.72\pm 0.005$ and $D=(4.84\pm 0.08)a^{2}V/\hbar $. In the
inset is shown a schematics of the system. As shown, the layers are fully
connected, and periodic boundary conditions are applied to the longitudinal
dimension.}
\end{figure}

\begin{figure}[tb]
\centering \leavevmode
\center{\epsfig{file=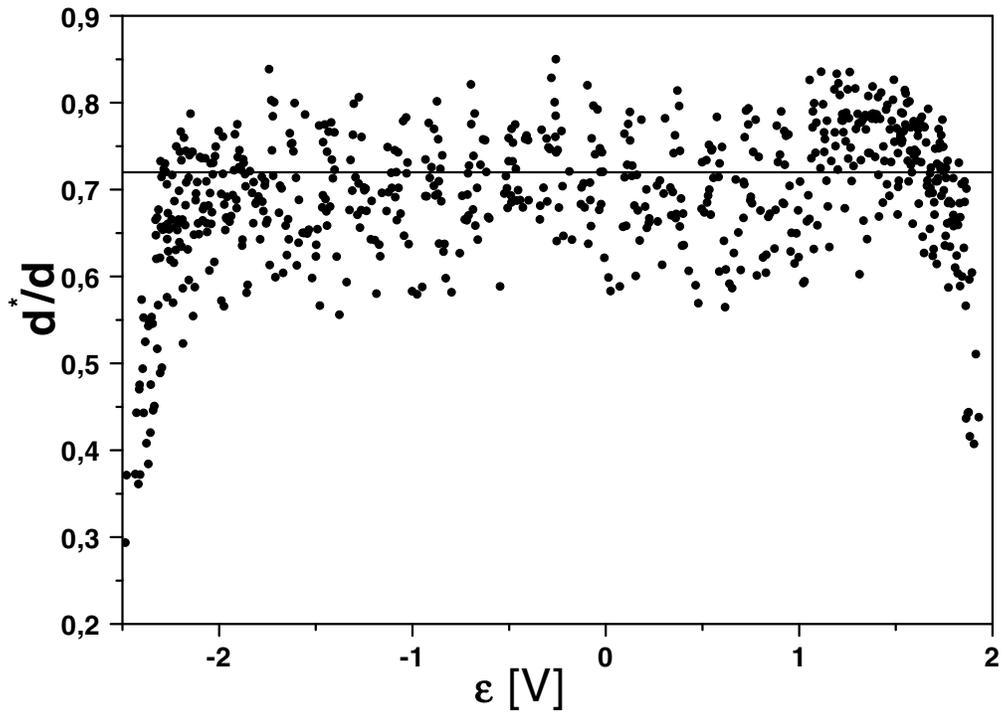, ,width=16cm,angle=270}}
\caption{Calculation of the effective dimension of every eigenstate of the
system by means of the inverse participation number (dots) compared to the
effective dimension obtained from the fitting of the dynamics of the system.
Except from very localized states in the band edges, agreement is quite
good. }
\end{figure}

\end{document}